\documentclass[a4paper]{article}

\usepackage{INTERSPEECH2018}
\usepackage{subfigure}
\usepackage{multirow}
\usepackage[colorlinks,linkcolor=black,anchorcolor=black,citecolor=black,urlcolor=black]{hyperref}
\title{Multi-task WaveNet: A Multi-task  Generative Model  for Statistical Parametric Speech Synthesis without Fundamental Frequency Conditions}
\name{Yu Gu, Yongguo Kang}
\address{
  Baidu Speech Department\\
  Baidu Technology Park, Beijing, 100193, China
 }
\email{guyu04@baidu.com, kangyongguo@baidu.com}

\begin{document}

\maketitle
\begin{abstract}
 This paper introduces an improved generative model for statistical parametric speech synthesis (SPSS) based on WaveNet under a multi-task learning framework.  Different from the original WaveNet model, the proposed Multi-task WaveNet employs the frame-level acoustic feature prediction as the secondary task and the external fundamental frequency  prediction model for the original WaveNet can be removed.
 Therefore the 
improved WaveNet  can generate high-quality speech waveforms only conditioned on linguistic features. Multi-task WaveNet can produce more natural and expressive speech by addressing the  pitch prediction error accumulation issue and possesses  more   succinct inference procedures than  the original WaveNet.
   Experimental results prove that the SPSS method proposed in this paper can achieve better performance than the  state-of-the-art approach utilizing  the original WaveNet in both objective and subjective preference tests.

\end{abstract}
\noindent\textbf{Index Terms}: WaveNet, multi-task learning, statistical parametric speech synthesis

\section{Introduction}
\label{sec:intro}
Text-to-speech (TTS) synthesis involves generating intelligible and natural  sounding synthetic speech
waveforms given the input text messages. At present, TTS synthesis technique is an indispensable basic component 
in various applications with speech interface such as car navigation systems, speech-to-speech translation, voice assistant and screen readers, etc. Thus the  demands and expectations for high-quality, high-naturalness, more stable and expressive speech waveform synthesis algorithms will be increasing more and more in future.

Statistical parametric speech synthesis (SPSS) \cite{zen2009statistical} is one of the mainstream speech synthesis techniques in which statistical models are employed to model the complex mapping relationship between the input linguistic information and the corresponding acoustic features. Decision-tree-clustered context-dependent hidden Markov model (HMM) speech synthesis \cite{hmmspss1} with single Gaussian state-output distribution has dominated
SPSS in the past decade. In  recent years, deep learning technology has been intensively studied and adopted in many speech generation tasks \cite{dl_tts}. Different kinds of stochastic neural
networks with various deep  structures have also been utilized as the acoustic models to replace the
decision trees and the probability density functions over acoustic features in conventional HMM-based
 SPSS methods. These models can model the relationship between input
features and acoustic features more accurately than conventional
HMMs and Gaussian mixture models and they can be used to model high-dimensional spectra
directly.
Unidirectional or bidirectional recurrent neural networks (RNNs) incorporating long short-term memory (LSTM) cells
 of inherently strong ability in
capturing long range temporal dependencies were also exploited on
speech synthesis systems to produce higher quality and
smoother speech trajectories than conventional deep neural
networks \cite{tts_lstm, ZenS15} .

Comparing with unit-selection speech synthesis, SPSS is more flexible and has the advantages on TTS tasks for a small-scale and mildly-curated speech corpus and  speech synthesis on mobile devices. The synthesized speech by SPSS is much more stable and  can effectively alleviate the
discontinuity which is a representative drawback of speech synthesis based on unit selection and waveform concatenation.
Nevertheless,  due to the limits of  some factors such as  vocoder quality, modeling accuracy and over-smoothing effect, the  quality and similarity  of generated speech from SPSS are still far from those of natural speech.
WaveNet \cite{wavenet} proposed recently by DeepMind is one of the state-of-the-art generative models in speech generation area. WaveNet and its variants with similar dilated convolutional network structures have achieved huge success on multiple  audio generation tasks besides speech synthesis such as speech enhancement \cite{wavenet_se}, voice conversion \cite{vcwavenet}, singing synthesis \cite{blaauw2017neural, wavenet_singing}, speech bandwidth extension \cite{gu2017waveform} and neural vocoders \cite{wavenet_vocoder, deepvoice2, deepvoice3}. Distinguished from conventional frame-based SPSS approaches, 
WaveNet can model the speech waveforms directly using dilated causal convolutional neural networks (CNNs) instead of vocoders.
It has been proved that WaveNet was capable of producing significantly more natural sounds than conventional SPSS approaches \cite{wavenet, deepvoice}, therefore WaveNet  is an  effective and promising solution to fill the gap of quality between natural and synthesized speech in SPSS.

Original WaveNets for the SPSS task were locally  conditioned on the logarithmic fundamental frequency
(log F0) values in addition to the linguistic features. The pitch or fundamental frequency information of speech is crucial for waveform generation because it represents the periodicity of speech waveforms. 
The absence of the log F0 condition information could 
degrade the naturalness of generated speech and lead in severe intonation mistakes \cite{wavenet}, therefore a different auxiliary model for predicting log F0 contours  from linguistic features was quite essential for waveform generation in WaveNet. However the involvement of an external F0 prediction model made the WaveNet inference procedures much more complicated. The error of the pitch determinations and the mistakes for voiced/unvoiced decisions in the F0 prediction model could also bring in inaccurate or wrong input condition information for WaveNet, which 
directly  alleviated the  naturalness and quality  of the synthesized speech, even though WaveNet model itself was accurate.  
In this paper, an improved WaveNet for SPSS under the multi-task learning framework is proposed to address these issues. The proposed Multi-task WaveNet can  get rid of the redundant F0 prediction model by deploying the F0 prediction as a secondary task. 
Both objective and subjective tests show that the proposed Multi-task WaveNet can 
outperform the original WaveNet and can obtain smaller objective distortion  and better subjective preference results.

This paper is organized as follows. Section \ref{sec:wavenet} gives a brief review of the model structure of WaveNet.
Section \ref{sec:mtlwavenet} introduces the proposed Multi-task WaveNet in this paper and the constructed SPSS system in detail. The experimental conditions and results are described in   Section \ref{sec:exp} and finally Section \ref{sec:con}  concludes this paper.
\begin{figure}[t]
\centering
\includegraphics[width=0.95\linewidth]{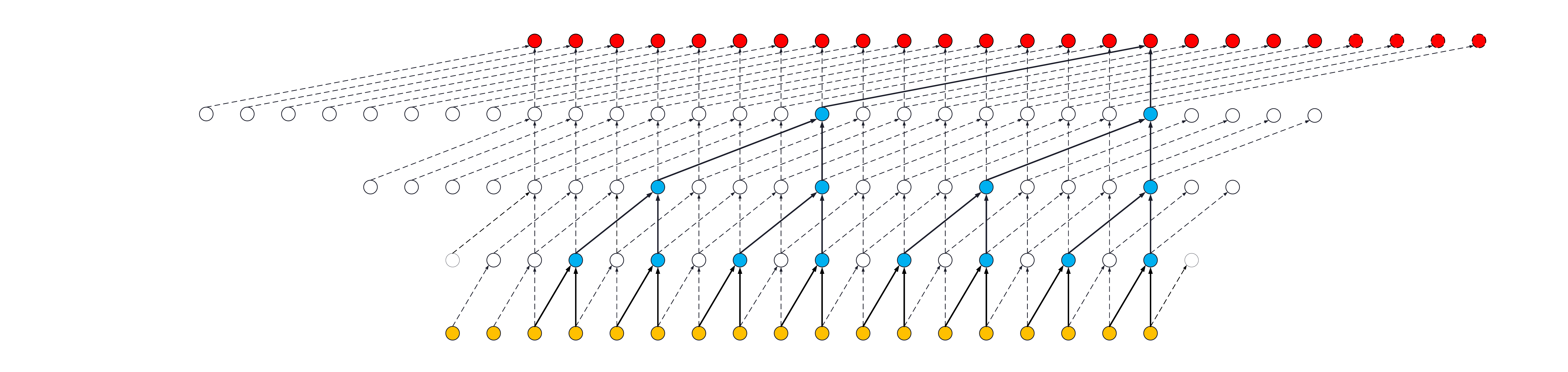}
\vspace{-0.3cm}
\caption{{Dilated causal convolutional network structures in WaveNet.}}
\label{fig:wavenet}
\end{figure}
\vspace{-0.3cm}
\section{WaveNet}
\label{sec:wavenet}
WaveNet \cite{wavenet} is a generative model which was proposed for TTS synthesis and other general audio generation tasks. WaveNet performed autoregressive speech sample generation  using an acoustic model with stacked dilated causal convolutional layers instead of depending on vocoders. The architectures of exploited dilated CNNs are illustrated
 on  Figure \ref{fig:wavenet}. 
 The causal convolutional layers have various dilation factors that allow their receptive field to grow exponentially in terms of the depths of networks as opposed to linearly, and can therefore cover the input history information from thousands of  timesteps ahead. 
  The dilated causal  CNN  can be regarded as a statistical model and the conditional  distribution of the output sample sequence  $\bm{x}=\lbrace x_{1}, x_{2}, \cdots\!, x_{T}\rbrace$ given the input local condition sequence $\bm{c}$ is  factorised as the product of conditional probabilities as follows:
\begin{equation}
\label{equ:one}
p\left(\bm{x}\mid\bm{c}\right)=\prod_{i=1}^{T}p\left(x_{i}\mid x_{i-N+1}, x_{i-N+2},\cdots\!, x_{i-1}, \bm{c}\right),
\end{equation}
where $N$ is the length of the receptive field.

Residual learning strategies  \cite{he2016deep} were also applied on the dilated CNNs in WaveNet to address the issues of training accuracy degradation and slow convergence.   Each convolutional layer in WaveNet is wrapped in  a residual block which contains gated activation units and two additional convolutional layers  towards the following and output layers respectively with convolution filters of size 1.
The residual and parameterized skip-connections are deployed throughout the network
to capacitate training  deeper networks and to accelerate convergence. The gated activation units with condition sequence $\bm{c}$ in $k$-th layer are expressed as:
\begin{align}
\bm{\hat{h}}_{k}=&\tanh(\bm{W}_{f,k}\ast\bm{h}_{k} +\bm{V}_{f,k}\ast\bm{c})  \ \odot  \notag\\
&\sigma(\bm{W}_{g,k}\ast\bm{h}_{k}+\bm{V}_{g,k}\ast\bm{c} ),
\end{align}
where  $f$ and $g$ denote the filter and gate parts respectively, $\sigma$ is the sigmoid non-linearity function, $\odot$ is the element-wise product and
$\ast$ is the convolution operator. The output layer is cascaded with a softmax layer and  thus the model can  describe the categorical distribution of waveform sample values encoded by $\mu$-law algorithm \cite{itu711}. 

\begin{figure}[t]
\setlength{\belowcaptionskip}{-0.2cm}
\centering
\includegraphics[width=\linewidth]{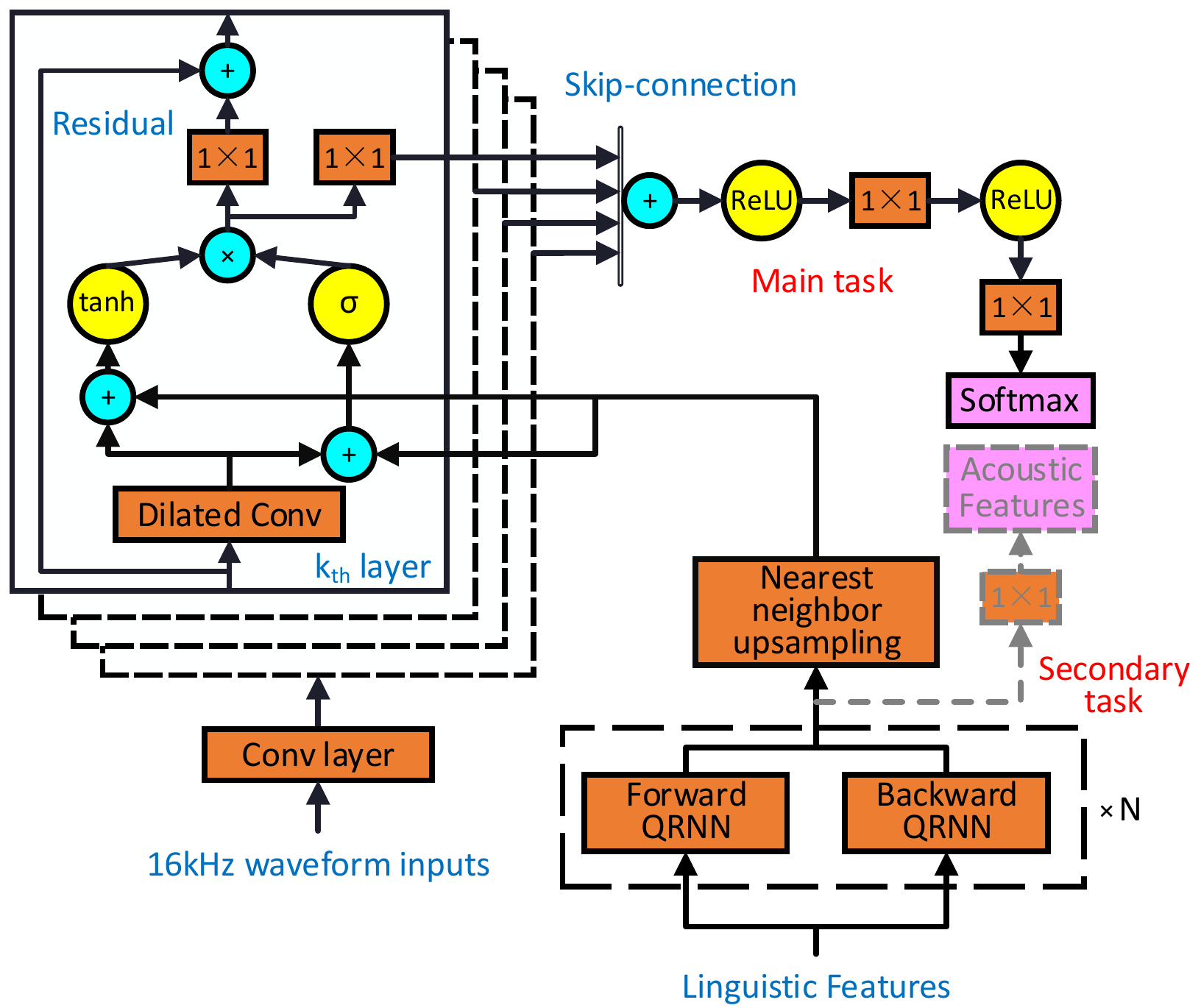}
\vspace{-0.3cm}
\caption{{Network structures of Multi-task WaveNet.}}
\label{fig:residual}
\end{figure}
\section{Multi-task WaveNet}
\label{sec:mtlwavenet}
\subsection{Multi-task learning}
Multi-task learning (MTL) \cite{caruana1998multitask} is a very  useful machine learning strategy which aims at improving model generalization ability and performance by  jointly learning several different but related tasks.
The main task usually shares a part of the representation with  
 other secondary tasks and the secondary tasks can contribute to 
 the model training of the corresponding main task by supplementing  information, transferring knowledge and increasing the amount of training data.
 MTL approaches can be easily deployed on stochastic neural networks by sharing  certain hidden layers cross different tasks.
 MTL has obtained great achievements on multiple speech signal processing  areas such as speech synthesis \cite{WuVWK15} and automatic speech recognition \cite{huang2013cross}. 
 
In this paper, the MTL strategy is also employed on WaveNet to assist the network training.
For the SPSS method based on the original WaveNet, the core task is to generate the speech samples in an autoregressive manner given the input linguistic features and log F0 values,
meanwhile the task of the conventional frame-based SPSS approach is to predict the frame-level acoustic features extracted by vocoders from the input linguistic features.
 These two different tasks share a part of the same inputs and the outputs of the two tasks are highly correlated,
  therefore the task of acoustic feature prediction can be treated as the secondary task for WaveNet. 
 As exhibited in Figure \ref{fig:residual}, the secondary task shares the same conditional network with WaveNet.
 The output acoustic features of the secondary task  consist of spectral features and fundamental frequency information, thus the conditional network of the trained Multi-task WaveNet possesses the representation capacity of 
 fundamental frequencies corresponding with the target speech waveforms.  
 Due to the pitch information complement from the secondary task, the linguistic features can be 
 directly utilized as the condition information instead of concatenating with log F0 values in  Multi-task WaveNet.
 The proposed Multi-task WaveNet owns several advantages over the original WaveNet. Firstly, the F0 prediction model is no longer needed, which will greatly simplify the inference procedures of WaveNet-based SPSS. Secondly, the inconsistency of using natural F0s for training and using predicted F0s for testing in the original WaveNet doesn't exist in the proposed model and the error accumulation problem from the F0 prediction model  is also tackled.
 Finally, the secondary task is also conducive to accelerating the convergence of WaveNet training.  

\subsection{Quasi-recurrent network condition}
Similar with the network architectures in Deep Voice \cite{deepvoice}, quasi-recurrent neural networks (QRNNs) \cite{qrnn}  with a stack of bidirectional RNN layers are employed in Multi-task WaveNet as the conditional networks to encode the linguistic input information.
A unidirectional QRNN layer with fo-pooling \cite{qrnn} is defined by the following equations:
\begin{align}
\label{equ:qrnn}
\hat{\bm{h}}&=\tanh(\bm{W}_{h}\ast\bm{x}+\bm{B}_{h}),\\
\bm{o}&=\sigma(\bm{W}_{o}\ast\bm{x}+\bm{B}_{o}),\\
\bm{f}&=\sigma(\bm{W}_{f}\ast\bm{x}+\bm{B}_{f}),\\
\bm{h}_{t}&=\bm{f}_{t}\cdot\bm{h}_{t\!-\!1}+(1-\bm{f}_{t})\cdot\hat{\bm{h}}_{t},\\
\bm{z}_{t}&=\bm{o}_{t}\cdot\bm{h}_{t},
\end{align}
where $\ast$ denotes the convolution operator. 
Bidirectional QRNN layer is computed by running two unidirectional QRNNs, one on the input sequence and one on
a reverse of the input sequence. Following the bidirectional QRNN layers, the encoded information is upsampled to the same time resolution with native audio frequency by repetition. QRNNs allow for parallel computation across the 
timestep dimension and have faster training and testing speeds than conventional RNNs. 
\subsection{SPSS using Multi-task WaveNet }
At the training stage of the SPSS method based on Multi-task WaveNet, a conventional decision tree-clustered context-dependent HMM-based SPSS model is trained to acquire the alignment information of the corpus.
And then a duration model based on a bidirectional RNN with stacked LSTM layers is trained using the obtained alignment information. The input linguistic features for WaveNet include some one-hot features for categorical linguistic contexts (e.g. phonemes identities, stress marks) and some numerical features for numerical linguistic contexts (e.g. the number of syllables in a word, the position of the current frame in the current phoneme). 
The output acoustic features for the secondary task including mel-cepstral coefficients (MCCs) and F0 values are extracted from natural speech by 
vocoders. Similar with the original WaveNet, all the input
and output waveforms are quantized to 256 discrete values using $\mu$-law  and one-hot coding is pursued on the quantized waveforms as the input waveform sequence. For the main task, the network
training is based on cross-entropy criterion to iteratively
improve the classification accuracy of output samples with
the target output sample sequences in training set.
For the secondary task, minimum mean squared error criterion is 
applied to estimate the MCCs, log F0 and voice/unvoice flag.

At the stage of inference, the phone-level linguistic features generated by front-end text analysis are  delivered to the duration prediction neural network to produce 
the duration information. Then combining with the duration information, WaveNet model can perform autoregressive speech sample generation conditioned on the frame-level linguistic features and the secondary task part as shown by the grey dashed line in Figure \ref{fig:residual} is abandoned at the speech synthesis phase.

\section{Experiments}
\label{sec:exp}
\subsection{Experimental conditions}
To evaluate the performance of the proposed Multi-task WaveNet,
 two  different scales of speech database pronounced by two Chinese female speakers respectively were used in the experiments. The large-scale corpus contained 16.2 hours of speech data, which was sufficient for unit-selection synthesis and was named as `` \emph{Corpus A} ''.
 The small-scale corpus named as ``\emph{Corpus B}'' only included 2.1 hours of speech data whose corresponding  transcripts were mainly designed for  car navigation. The sampling rate of the speech data was 16 kHz and 50 test sentences those were not present or similar to those in the training set were used as the test set to
measure the performance of different speech synthesis systems  for evaluation.
For the main task,  the WaveNet model consisted of 40 layers, which were grouped into 4 dilated residual block stacks of 10 layers. In every stack,  the dilation rate exponentially increased by a factor of 2 in every layers which started
with rate 1  and ended with the maximum dilation of 512.  For the secondary task, 
 25-dimensional MCCs were extracted  from the smoothed spectral envelopes
 obtained by STRAIGHT analysis \cite{straight} and the F0 values
were extracted by RAPT algorithm \cite{rapt}. 

The following
five speech synthesis systems were established for comparison.

\begin{itemize}
\setlength{\itemsep}{0pt}
\item \emph{\textbf{LSTM}}: The SPSS system using  bidirectional RNNs with stacked LSTM layers as duration and acoustic models;
\item \emph{\textbf{Concatenative}}: The HMM-driven unit selection concatenative speech synthesis;
\item \emph{\textbf{WaveNet}}: The original WaveNet SPSS system with  F0s and linguistic features as WaveNet conditions;
\item \emph{\textbf{WaveNet-lin}}: The original WaveNet SPSS  system only conditioned on linguistic features;
\item \emph{\textbf{MTL-WaveNet}}: The proposed SPSS system using Multi-task WaveNet.
\end{itemize}

\begin{table}[t]
\vspace{-0.2cm}
 \centering
 \renewcommand{\arraystretch}{1.1}
  \caption{Comparison of distortion between acoustic features of natural speech and synthesized speech from different systems.
    V/UV  means frame-level voiced/unvoiced error. BAP and Corr. represents the
BAP prediction error and correlation coefficients respectively.}
  
 \vspace{-0.4mm}
\subtable[Results for \emph{Corpus A}]{
     
        \setlength{\tabcolsep}{1.5mm}{
          \begin{tabular}{| c |ccccc|}
            \hline
             \multirow{2}{*}{System} & MCD &  BAP & F0 RMSE  &F0& V/UV  \\
              &  (dB) & (dB) & (Hz)&  Corr.&(\%)\\
            \hline \hline
            \emph{\textbf{LSTM}}  & 2.265 & 2.131 & 30.485 &0.860&5.264 \\
           \emph{\textbf{WaveNet-lin}} & 1.524 & 2.798 & 39.413 & 0.756 &4.745\\
            \emph{\textbf{WaveNet}}  & 1.548 & 2.787 & 32.206& 0.845&4.796\\
           \emph{\textbf{MTL-WaveNet}}& \textbf{1.519} & 2.712 & \textbf{22.396}  & \textbf{0.922}&\textbf{4.298}\\
            \hline
          \end{tabular} }    
          }
     \subtable[Results for \emph{Corpus B}]{
  
        \setlength{\tabcolsep}{1.5mm}{
          \begin{tabular}{| c |ccccc|}
            \hline
            \multirow{2}{*}{System} & MCD &  BAP & F0 RMSE  & F0 & V/UV  \\
              &  (dB) & (dB) & (Hz)& Corr. &(\%)\\
            \hline \hline
            \emph{\textbf{LSTM}}  & 1.641 & 2.053 & 37.021 &0.787&4.328 \\
           \emph{\textbf{WaveNet-lin}} & 1.515 & 2.683 & 41.745 & 0.756 &3.588\\
            \emph{\textbf{WaveNet}}  & 1.500 & 2.681 & 37.790& 0.775& 3.592\\
           \emph{\textbf{MTL-WaveNet}}& \textbf{1.481} & 2.653 & \textbf{33.801}  & \textbf{0.821}&3.682\\
            \hline
          \end{tabular} }    
          }  
           \label{tab:obj}
      \end{table}
   \vspace{-0.2cm}
\setlength{\abovecaptionskip}{0.2cm}
\setlength{\belowcaptionskip}{-0.2cm}
 
\subsection{Objective evaluation}
Objective tests were conducted to evaluate different synthesis systems. Because the duration of synthetic speech in the \textbf{\emph{Concatenative}} system was intractable to adjust, 
the \textbf{\emph{Concatenative}} system was 
excluded from the objective evaluations and  the other SPSS systems remained the same ground-truth duration as the target natural speech for the convenience of comparison.
Mel-cepstral distortion (MCD),  distortion of band aperiodicities (BAP), voiced/uvoiced prediction error,  root-mean-square error (RMSE) and correlation coefficients of F0 values on a linear scale  between natural speech and synthesized speech by different SPSS systems are presented in Table \ref{tab:obj}. It is worth mentioning that  for the \emph{\textbf{WaveNet}}, \emph{\textbf{WaveNet-lin}} and \emph{\textbf{MTL-WaveNet}} systems, the compared acoustic features  were re-extracted from
the generated waveforms, while those were directly the model outputs for the  \emph{\textbf{LSTM}} system. 

The objective results for both two speech databases show that the synthesized speech from the \emph{\textbf{MTL-WaveNet}} system can acquire more  accurate F0 values and smaller spectral distortion.  The F0 RMSE of the \emph{\textbf{WaveNet-lin}} system  is the largest among all the systems, which indicates the log F0 conditional information is quite essential for the original WaveNet. 
 The F0 prediction error of the \emph{\textbf{WaveNet}} system is also larger than the conventional \emph{\textbf{LSTM}} system which means the inaccuracy of the F0 prediction model can be accumulated into the waveform generation step of WaveNet and further impact the synthesized speech quality in the \emph{\textbf{WaveNet}} system.

\subsection{Subjective evaluation}
\begin{figure}[t]
\centering
\subfigure[Results for \emph{Corpus A}. The $p$-values of $t$-test in these comparisons are $5.6\times10^{-9}$, $8.8\times10^{-7}$ and $0.69$ respectively.]{
\includegraphics[width=0.95\linewidth]{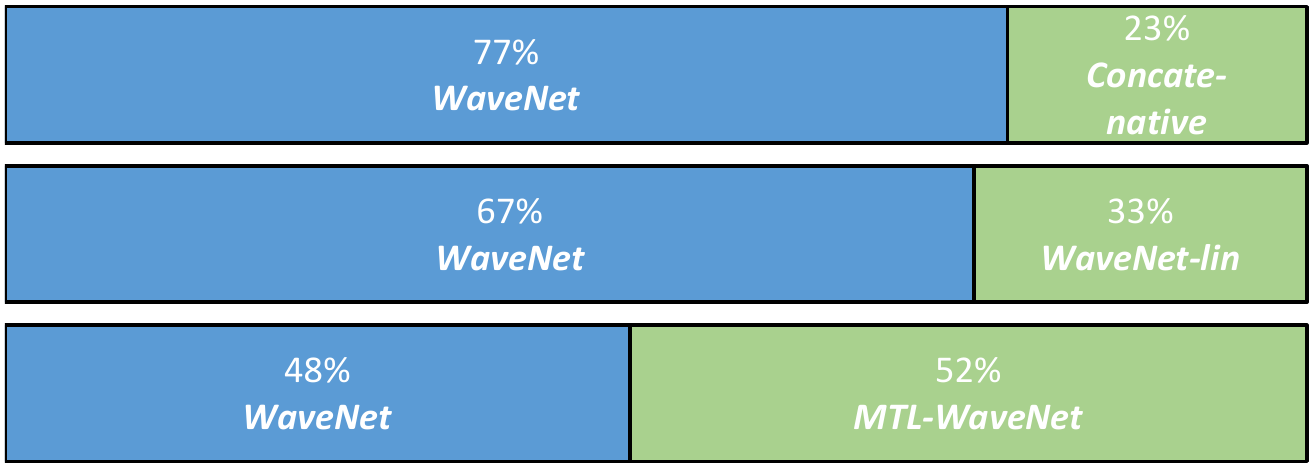}}
\subfigure[Results for \emph{Corpus B}. The $p$-values of $t$-test in these comparisons are $9.5\times10^{-19}$, $5.6\times10^{-9}$ and $0.43$ respectively.]{
\includegraphics[width=0.95\linewidth]{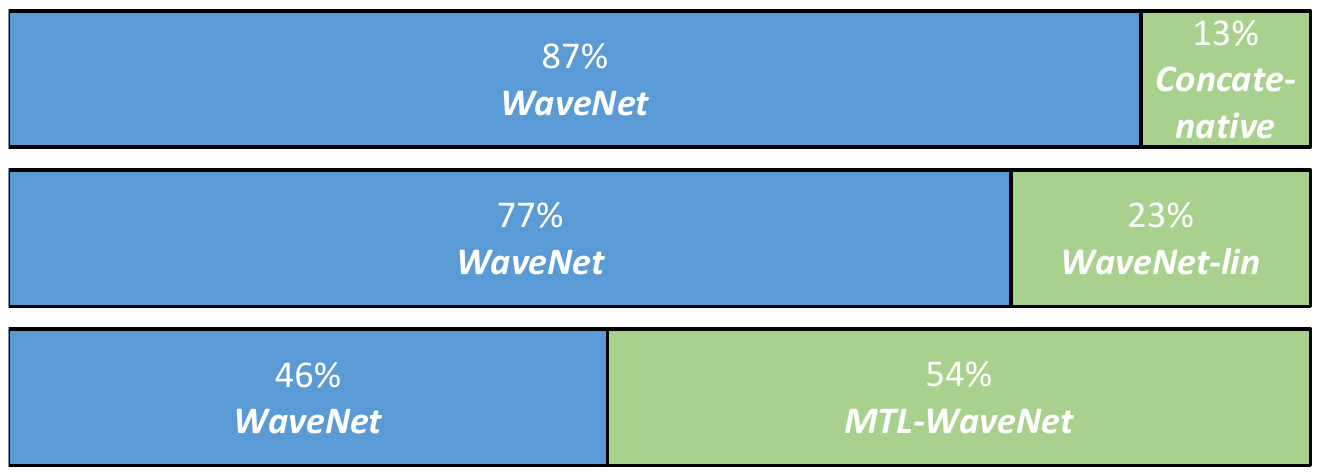}}
\vspace{-0.3cm}
\caption{{Preference test scores among different TTS systems. }}
\vspace{-0.3cm}
\label{fig:pref}
\end{figure}

Several preference tests were performed to  assess the subjective perceptual quality of the speech generated from different speech synthesis  systems. In each  preference test,  20 test utterances randomly selected from the test set were synthesized by two different systems and evaluated in random order by five listeners.
Because the speech quality of the \emph{\textbf{LSTM}} system was far from  other systems, we only compared the preference performances among the \emph{\textbf{Concatenative}}, \emph{\textbf{WaveNet-lin}}, \emph{\textbf{WaveNet}} and \emph{\textbf{MTL-WaveNet}} systems.
 The listeners were asked to choose their preference for each  pairwise utterances in terms of speech quality.\footnote{Examples of synthesized speech by different systems are available at \url{https://ttsdemos.github.io}.} 
 The preference scores of listening tests conducted in two databases of different scales are exhibited in  Figure \ref{fig:pref}  respectively with the $p$-values from $t$-test.
The comparisons of the conventional  \emph{\textbf{Concatenative}} system and the  \emph{\textbf{WaveNet}}  systems demonstrate the  WaveNet based method using waveform modeling and causal dilated CNNs can successfully improve the quality of synthesized speech. Although the candidate and selected units for the  \emph{\textbf{Concatenative}} system are natural speech segments in the corpus, the discontinuity of the 
synthesized speech in unit-selection synthesis seriously degrades the perceptual quality and naturalness. The gap between the \emph{\textbf{Concatenative}} and \emph{\textbf{WaveNet}} systems for \emph{Corpus B} is much larger than that for \emph{Corpus A},
which is because the data volume of \emph{Corpus B} is not sufficient for building a unit-selection synthesis system and there are more bad cases than \emph{Corpus A}. The data size can also affect the synthesized speech quality for the approaches based on WaveNets, while the 
differences are not as obvious as those for unit-selection synthesis. The comparisons between the   \emph{\textbf{WaveNet-lin}} and  \emph{\textbf{WaveNet}} systems can also prove the importance of using log F0 as a part of condition information in original WaveNet SPSS methods.   We find that WaveNet only
conditioned on linguistic features can synthesize natural  waveforms for the most part of one utterance 
but sometimes it has unnatural phones and  syllables   by pronouncing wrong and strange tones.
The superiority of the \emph{\textbf{MTL-WaveNet}}  system over the \emph{\textbf{WaveNet}} system on preference scores indicates  the effectiveness of utilizing the acoustic feature prediction as the secondary task.
The improvement of preference scores for the \emph{\textbf{MTL-WaveNet}} is less significant than those of objective tests. In fact, 
the generated speech from the \emph{\textbf{WaveNet}} system is generally good enough and the advantages of the proposed Multi-task WaveNet 
 are embodied on some certain speech segments and some speech details which are easily ignored by listeners.

\section{Conclusions}
\label{sec:con}
In this paper, we propose an improved WaveNet model for SPSS exploring the multi-task learning strategy. The frame-level acoustic 
feature prediction is introduced as the auxiliary secondary task to supplement  the requisite pitch information. Comparing with the original WaveNet 
based SPSS approach, the proposed Multi-task WaveNet can get rid of the redundant F0 prediction model and increase the inference efficiency. This model
 can also solve the  pitch prediction error accumulation problems.
 Both objective and subjective evaluation results show that the SPSS method proposed in this paper have the advantages over the original WaveNet. To achieve more natural prosody and more expressive speech, some end-to-end speech synthesis models will be tried to further replace the duration models for WaveNets in our future work. 
We will deploy such multi-task learning structure on Parallel WaveNet to further accelerate the inference speed and apply the proposed algorithms on online products.

\section{Acknowledgements}

The authors thank  Xian Li for his early constructive contribution to this work, as well as Zhenyu Wang, Hao Li, Xiaohui Sun, Feiya Li of Baidu Speech Department for their assistance  with data preprocessing  and running subjective evaluations. The authors also thank  Wei Ping and Kainan Peng of  Baidu Silicon Valley AI Lab for their helpful  discussions and advices.
\bibliographystyle{IEEEtran}

\bibliography{mybib}

\end{document}